\documentclass[
    ,final            
  ]
  {aipproc}

\layoutstyle{6x9}

\def\ga{\mathrel{\mathchoice {\vcenter{\offinterlineskip\halign{\hfil
$\displaystyle##$\hfil\cr>\cr\sim\cr}}}
{\vcenter{\offinterlineskip\halign{\hfil$\textstyle##$\hfil\cr>\cr\sim\cr}}}
{\vcenter{\offinterlineskip\halign{\hfil$\scriptstyle##$\hfil\cr>\cr\sim\cr}}}
{\vcenter{\offinterlineskip\halign{\hfil$\scriptscriptstyle##$\hfil\cr>\cr
\sim\cr}}}}}

\begin{document}

\title{Ultra-High Energy Cosmic Rays:\\Results and Prospects}


\author{Karl-Heinz Kampert}{
  address={University of Wuppertal, Department of Physics, Gau{\ss}strasse 20, D-42119 Wuppertal}\thanks{email: kampert@uni-wuppertal.de}
}

\begin{abstract}
Observations of cosmic rays have been improved at all energies, both in terms of higher statistics and reduced systematics.
As a result, the all particle cosmic ray energy spectrum starts to exhibit more structures than could be seen previously. Most importantly, a second knee in the cosmic ray spectrum -- dominated by heavy primaries -- is reported just below $10^{17}$\,eV. The light component, on the other hand, exhibits an ankle like feature above $10^{17}$\,eV and starts to dominate the flux at the ankle. The key question at the highest energies is about the origin of the flux suppression observed at energies above $5\cdot10^{19}$\,eV. Is this the long awaited GZK-effect or the exhaustion of sources? The key to answering this question is again given by the still largely unknown mass composition at the highest energies. Data from different observatories don't quite agree and common efforts have been started to settle that question. The high level of isotropy observed even at the highest energies starts to challenge a proton dominated composition if extragalactic (EG) magnetic fields are on the order of a few nG or more. We shall discuss the experimental and theoretical progress in the field and the prospects for the next decade.
\end{abstract}

\maketitle


\section{Introduction}

The Texas-Symposium 2012 completes a series of conferences at which the discovery of cosmic rays (CR) a hundred years ago by Franz Viktor Hess has been commemorated. Less known is a breakthrough made 50 years ago by John Linsley: in 1962 he reported the first observation of a primary CR particle with an energy exceeding $10^{20}$\,eV \cite{Linsley:1963uo}. This event remains one of the most energetic CRs ever recorded and Linsley was extremely lucky to observe such an event after just three years of data taking of the 2\,km$^2$ large Volcano Ranch air shower array. The cosmic microwave background (CMB) radiation, discovered three years later, immediately led to the prediction of a flux suppression either due to photo-pion production by protons propagating through the CMB at energies above $5\cdot10^{19}$\,eV or due to photodisintegration of nuclei at about the same threshold energy. This process is known as GZK-effect, predicted independently by Greisen and by Zatsepin and Kuz'min \cite{greisen-66,zatsepin-66}. This has been the only firm prediction about a structure in the CR energy spectrum \cite{Kampert:2012bz} and it has taken nearly 50 years to observe such a feature in data \cite{Abbasi-08,Abraham-08c}. However, can we be sure about having observed the GZK-effect? The spectral feature may equally be caused by the exhaustion of nearby sources or by a mixture of both as will be discussed below.

At lower energies, Kulikov and Khristiansen \cite{Kulikov:1959vq} in 1958 reported a structure in the shower size spectrum which became known as the ``knee'' in the CR spectrum. Observations made by KASCADE \cite{KASCADE-05} and other air shower experiments showed that the mean mass increases above the knee, indicating that the knee marks the maximum acceleration energy of the most abundant Galactic sources. This led to speculations that heavy primaries would experience the same limitation of particle acceleration and a second, so-called ``Iron-knee'', would be observed around $E_{\rm knee}^{\rm Fe} \sim 26\times E_{\rm knee}^p$. Such an observation has now been reported and it may mark the end of the Galactic CR spectrum.

Large- and small-scale anisotropies in the arrival directions have been reported at TeV energies and now reach to beyond a PeV. This has come as a surprise and its interpretation starts to result in a better understanding and modeling of CR propagation within our Galaxy and about the transition from Galactic to EG CRs. At the highest energies, only upper limits on large scale anisotropies have been reported so far but, due to limited statistics, the amplitudes cannot be probed down to the levels observed below the knee region. Instead, a weak correlation of the highest energy particles to the position of nearby AGN has been reported by Auger \cite{Abraham-07e}. This gave support to the picture that some fraction of the highest energy CRs results from sources within about 200 Mpc distance.

Much progress has also been made in particle acceleration emphasizing the importance of non-linear effects in diffuse shock acceleration (DSA) with magnetic field amplification due to CR current driven instabilities. These effects may not only substantially increase the maximum energy reachable in CR accelerators but may also reduce the time scales required for the acceleration process. 
While Galactic CRs are believed to originate from supernova remnants (SNR), those at the highest energies are thought to originate in the lobes of Radio Galaxies (RG) if they are large and luminous enough and, again, a substantial energy is contained in the turbulent component of the magnetic field.

Thus, it is fair so say that enormous progress has been made in CR physics particularly in recent years, both in observations and in accompanying theory. However, despite such advancements, the key questions about the CR origin and acceleration remain open even 100 years after their discovery. This paper aims to address some key topics in the field.

\section{The Cosmic Ray Energy Spectrum}

\begin{figure}
\includegraphics[width=.55\textwidth]{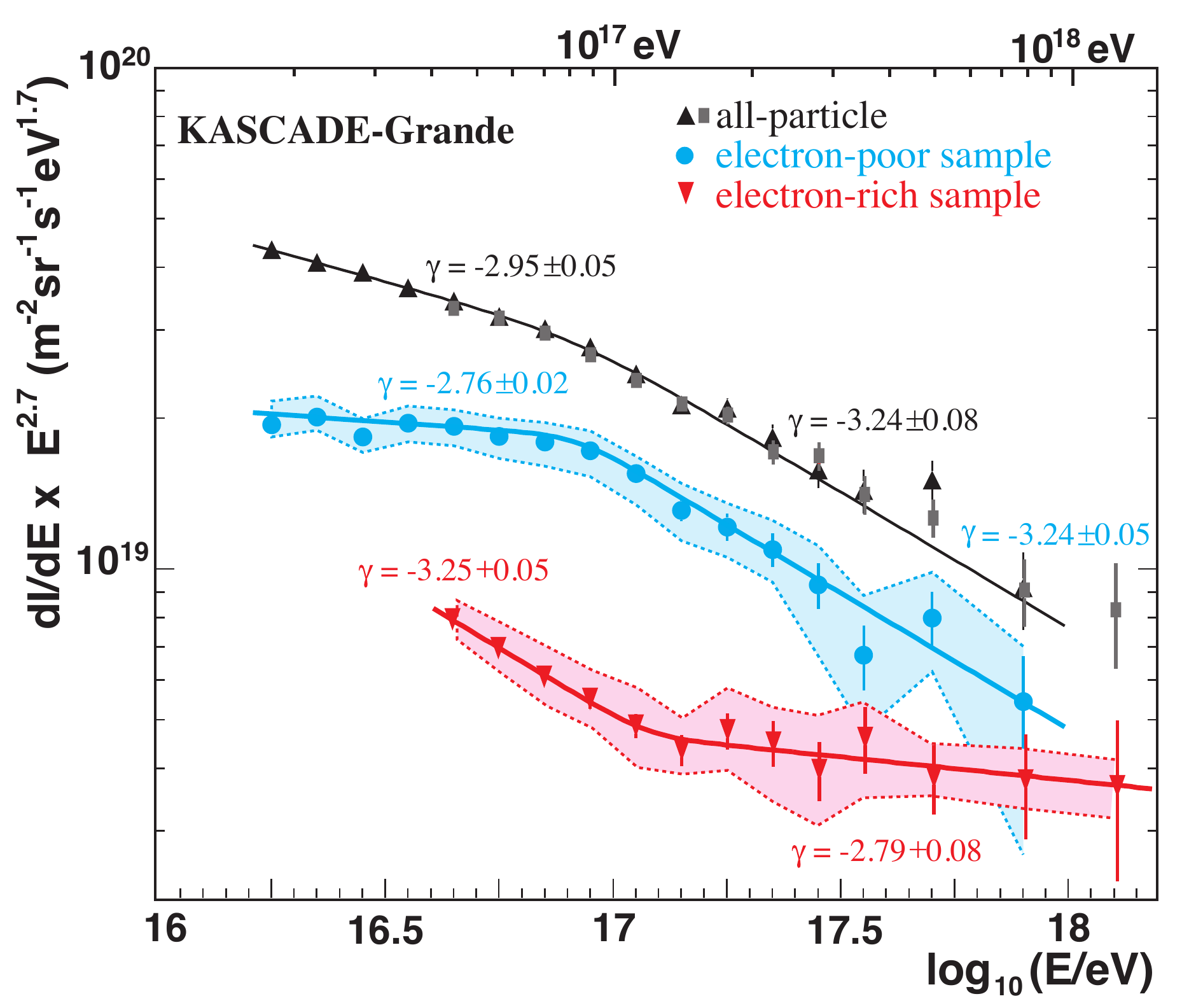}
\caption{All-particle, electron-poor, and electron rich energy spectra from KASCADE-Grande. The all-particle (black triangles; 105,000 events) and heavy enriched spectrum (blue circles; 52,000 events) is taken from \cite{Apel:2011bx} and the all-particle (grey squares) and light primary spectrum (red triangles; 6,300 events) results from a larger data set and employs stronger cuts to the light-component to select essentially p+He primaries \cite{Apel:2013-ankle}. The bands indicate systematic uncertainties resulting mostly from hadronic interaction models. The heavy enriched data sample exhibits a knee at $10^{16.9}$\,eV with a statistical significance of $3.5\sigma$ while the ankle-like feature in the light component is found at $10^{17.1}$\,eV with a significance of $5.8\sigma$. \label{fig:kg-spectra}}
\end{figure}

Recent progress in the knee-to-ankle energy range has been driven mostly by KASCADE-Grande with Tunka and IceTop ramping up and providing more data with high statistics and good resolution. Using complex 2-dimensional unfolding techniques to the electron vs muon numbers measured on shower-by-shower basis by the KASCADE air shower experiment, the mean mass was shown to become heavier above the knee energy with the energy spectra of primary mass groups supporting a scaling with rigidity according to $E_{\rm knee}^Z \simeq Z\times 3\cdot 10^{15}$\,eV \cite{KASCADE-05}, such as was suggested long time ago by Peters \cite{Peters:1961uq}. This observation, supported by other experiments, has renewed the question about the existence of a Fe-like knee at about $10^{17}$\,eV. Such a structure has been reported very recently and is shown in Fig.\,\ref{fig:kg-spectra}. The significance of the second knee at $E \simeq 80$\,PeV in the all-particle energy spectrum of KASCADE-Grande is just above $2\sigma$ but increases to $3.5 \sigma$ for the electron poor (heavy) sample \cite{Apel:2011bx}. Similarly, IceTop data \cite{Abbasi:2013id} show an indication of a flattening above 22 PeV, i.e.\ in the energy range between the two knees. Another very interesting recent result by KASCADE-Grande is reported in \cite{Apel:2013-ankle}. Using a larger data set and applying stronger cuts to electron-rich showers than were applied in \cite{Apel:2011bx}, to accept essentially only p+He primaries, there is an ankle-like feature at $10^{17.1}$\,eV with a significance of $5.8\sigma$.

Obviously, the CR energy spectrum, once measured with high precision, exhibits much more structure and information than just the knee energy and the indices of an apparent power-law like spectrum below and above. The observation of the `Fe-knee' and 'p-ankle' (with ``Fe'' and ``p'' meant as synonym for ``heavy'' and ``light'' primaries, respectively) is a remarkable achievement. The Fe-knee at $8\cdot 10^{16}$\,eV supports the picture of a rigidity scaling -- also named the ``Peters cycle''  \cite{Peters:1961uq} -- in the knee energy range and the p-ankle $E\simeq 1.2\cdot 10^{17}$\,eV has in fact been expected because of the steep fall-off of the p-component at the knee \cite{KASCADE-05} and the p-like composition at the ankle (see next section). Thus, the p-ankle would either mark the transition from Galactic to EG sources or the onset of a new high energy (Galactic) source population (see e.g.\ \cite{Jokipii:1987uv,Biermann:2012hx}).

\begin{figure}
\includegraphics[width=.9\textwidth]{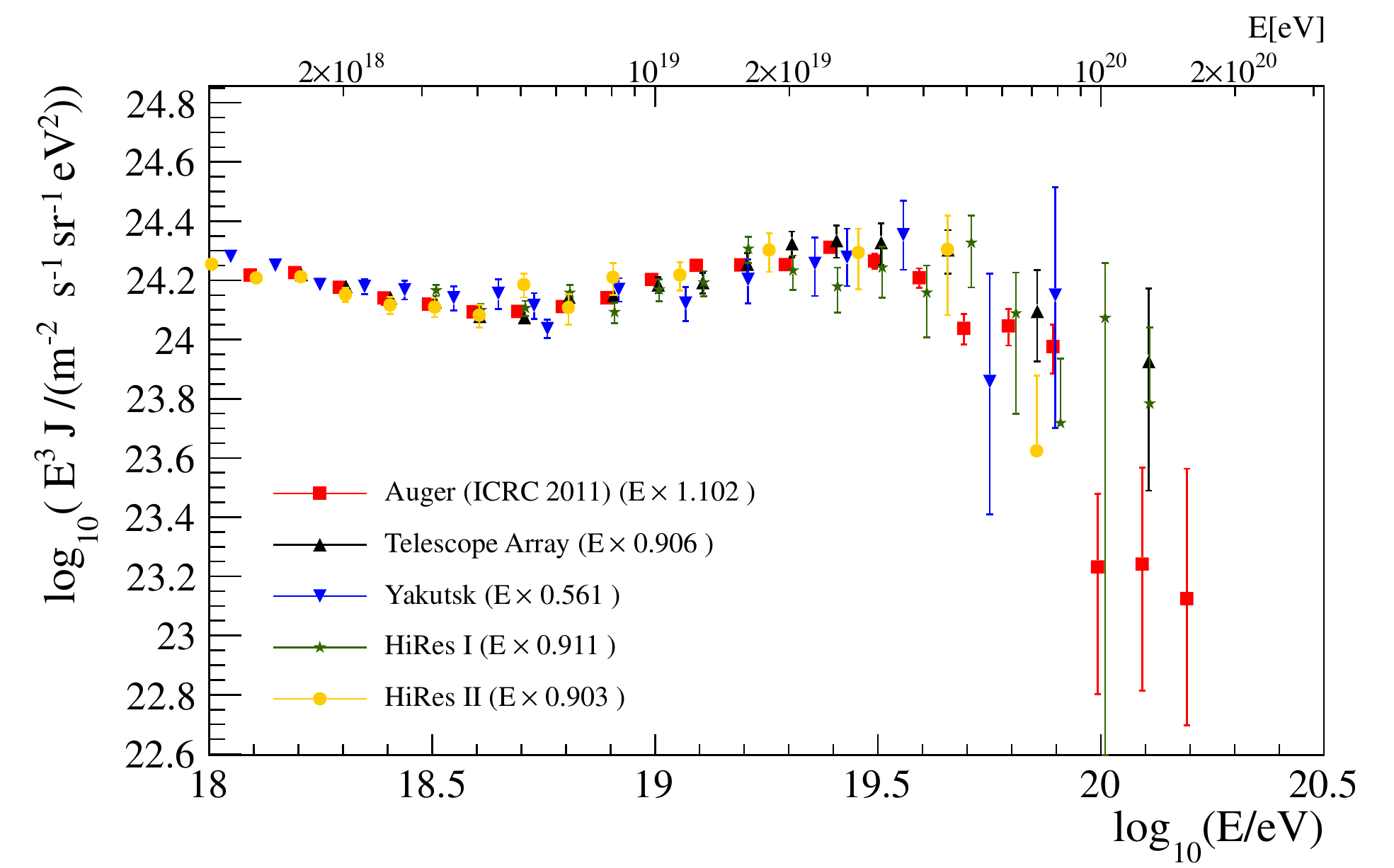}
\caption{Compilation of cosmic ray energy spectra, with the flux multiplied by $E^3$, published by Auger (combined Hybrid/SD), TA SD, Yakutsk SD, HiRes I, and HiRes II after energy-rescaling as shown in the figure has been applied. The reference spectrum is the average of those from Auger and TA. From \cite{E-WG} where also references to the respective data sets can be found.\label{fig:spectra2}}
\end{figure}

At the highest energies, from the ankle to beyond $10^{20}$\,eV, the Pierre Auger Observatory \cite{auger-04,Abraham-FD-10} is the flagship in the field with an accumulated exposure of about 30\,000\,km$^2$\,sr\,yr. The Telescope Array \cite{AbuZayyad:2012tr}, due to a later start and its more than 4 times smaller area, has collected about 10 times less events.  A detailed comparison of the energy spectra of various observatories is presented in Fig.\,\ref{fig:spectra2}. 
As discussed in great detail in \cite{E-WG}, it is found that the energy spectra determined by the larger experiments are consistent in normalization and shape after energy scaling factors, as shown in Fig.\,\ref{fig:spectra2}, are applied. Those scaling factors are within systematic uncertainties in the energy scale quoted by the experiments. This is quite remarkable and demonstrates how well the data are understood. Nevertheless, cross-checks of photometric calibrations and atmospheric corrections have been started and as a next step, common models (e.g.\ fluorescence yield) should be used where possible. The data in Fig.\,\ref{fig:spectra2} clearly exhibit the ankle at $\sim 4\cdot10^{18}$\,eV and a flux suppression above $\sim 4\cdot10^{19}$\,eV. The flux suppression at the highest energies is in accordance with the long-awaited GZK-effect \cite{Abbasi-08,Abraham-08c}. However, as discussed below, the data of the Auger observatory suggest that the maximum energy of nearby sources or the source population is seen, instead.

\section{Cosmic Ray Composition and Interaction Models}

Obviously the energy spectra by itself, despite their high level of precision reached, do not allow one to conclude about the origin of the spectral structures and thereby about the origin of CRs in different energy regions. Additional key information is obtained from the mass composition of CRs. Unfortunately, the measurement of primary masses is the most difficult task in air shower physics as such measurements rely on comparisons of data to EAS simulations with the latter serving as reference \cite{Kampert:2012hg}. EAS simulations, however, are subject to uncertainties mostly because hadronic interaction models need to be employed at energy ranges much beyond those accessible to man-made particle accelerators. Therefore, the advent of LHC data, particularly those measured in the extreme forward region of the collisions, is of great importance to CR and EAS physics and have been awaited with great interest \cite{Kampert-ISVHECRI12}. Remarkably, interaction models employed in EAS simulations provide a somewhat better description of global observables (multiplicities, $p_\perp$-distributions, forward and transverse energy flow, etc.) than typical tunes of HEP models, such as PYTHIA or PHOJET \cite{dEnterria:2011kga}. This demonstrates once more that the CR community has taken great care in extrapolating models to the highest energies. Moreover, as demonstrated e.g.\ in \cite{PAbreuetal:2012vi}, CR data provide important information about particle physics at centre-of-mass energies ten or more times higher than is accessible at LHC. The $pp$-inelastic cross section extracted from data of the Auger Observatory supports only a modest rise of the inelastic cross section with energy \cite{PAbreuetal:2012vi}.

\begin{figure}
\includegraphics[width=.56\textwidth]{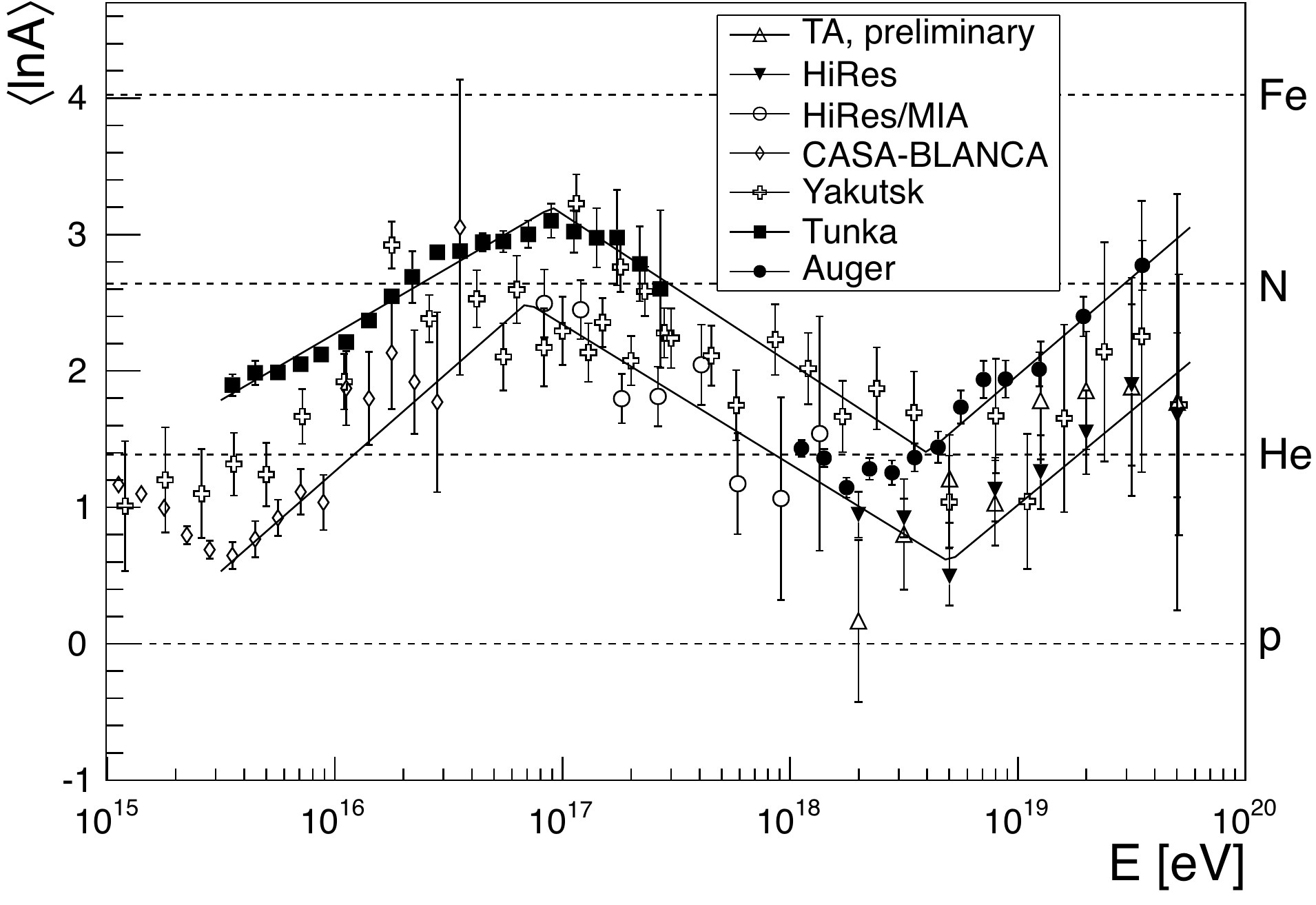}
\includegraphics[width=.44\textwidth]{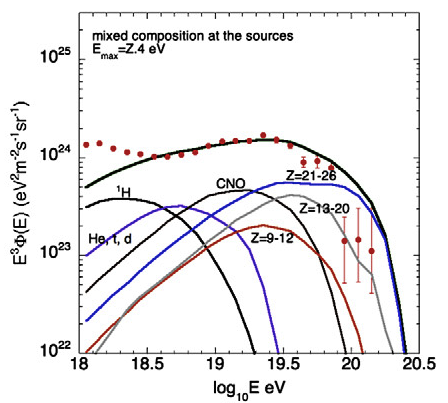}
\caption{Left: Average logarithmic mass of cosmic rays as a function of energy derived from $X_{\rm max}$ measurements with optical detectors for the EPOS 1.99 interaction model. Lines are estimates of the experimental systematics, i.e.\ upper and lower boundaries of the data presented \cite{Kampert:2012hg}. Right: Propagated CR spectrum assuming a mixed composition similar to the Galactic one with a maximum energy at the sources of $E_{\rm max}(Z) = Z\times4\cdot10^{18}$\,eV and a spectral index $\beta = 1.6$ \cite{Allard:2011ul}.\label{fig:compos}}
\end{figure}

A careful analysis of composition data from various experiments has been presented in \cite{Kampert:2012hg} with exemplary results depicted in Fig.\,\ref{fig:compos} (left). These data complement those of the energy spectrum in a remarkable way. As can be seen, the breaks in the energy spectrum coincide with the turning points of changes in the composition: the mean mass becomes increasingly heavier above the knee, reaches a maximum at the 2$^{\rm nd}$ knee, another minimum at the ankle before it starts to rise again towards the highest energies. Different interaction models provide the same answer concerning changes in the composition but differ by their absolute values of $\langle \ln A \rangle$. It should also be noted that the suggested increase in the mean mass at the highest energies is not without dispute. It has been looked at in great detail in \cite{compos-WG}.
At ultra-high energies, the Auger data suggest a larger $\langle \ln A \rangle$ than all other experiments. 
TA and Yakutsk are consistent within systematic uncertainties with Auger data while HiRes is compatible with Auger only at energies below $10^{18.5}$\,eV when using QGSJet-II. When using the SIBYLL model, Auger and HiRes become compatible within a larger energy range \cite{compos-WG}.

The importance of measuring the composition up to the highest energy cannot be overstated as it will be the key to answering the question about the origin of the GZK-like flux suppression. The same mechanism of limiting source energy that appears to cause the increasingly heavy above the knee may work also for EG-CRs above the ankle. Thereby, the break at $\sim 4\cdot10^{19}$\,eV may mark the maximum energy of nearby EG CR-accelerators, rather than the GZK-effect. This is shown in Fig.\,\ref{fig:compos} (right), where propagated CR spectra are shown for a maximum energy at the source of $E_{\rm max}(Z) = Z\times4\cdot10^{18}$\,eV and assuming a hard spectral source index of $\beta = 1.6$ \cite{Allard:2011ul}. Clearly, such a -- in view of the hard spectral index -- more exotic scenario provides a good description of the energy spectrum. Moreover, other than the GZK-like interpretation, it also describes the $\langle X_{\rm max} \rangle$ {\em and} the fluctuation RMS($X_{\rm max}$) of the Pierre Auger Observatory \cite{Abraham-xmax-10}.

A mixture of light and intermediate/heavy primaries at the highest energies may also explain the low level of directional correlations to nearby AGN. Enhancements, presently foreseen by the Auger Collaboration will address this issue.

Two models about the putative transition from Galactic- to EG-CRs have received much attention: In the classical `ankle model' the transition is assumed to occur at the ankle. In this model, Galactic CRs above the $2^{\rm nd}$ knee are dominated by heavy primaries before protons of EG origin start to take over and to dominate at the ankle. In the dip-model \cite{berezinsky-88}, on the other hand, the transition occurs already at the $2^{\rm nd}$ knee and is characterized by a sharp change of the composition from Galactic iron to EG protons while the ankle is due to $e^+ e^-$ production of protons in the CMB. A third, `mixed composition', model has been suggested more recently \cite{Aloisio:2012tx} in which EG-CRs taking over are not considered being protons but an EG mixed CR composition. Clearly, the dip-model requires a proton dominated composition essentially at all energies starting somewhat above the $2^{\rm nd}$ knee. The answer may be difficult to determine based on $\langle X_{\rm max} \rangle$ or $\langle \ln A \rangle$ alone. A much better quantity would again be the RMS of these quantities, such as studied at higher energies in \cite{Abraham-xmax-10}. A rather abrupt change of composition as required by the dip-model near the $2^{\rm nd}$ knee vs a smooth change of composition as expected near the ankle in the ankle model, should become distinguishable by the RMS$(X_{\rm max})$-values already in the very near future. This has been a prime motivation for the HEAT and TALE extensions of Auger and TA, respectively.

\section{Anisotropies at Different Energies and Angular Scales}

\begin{figure}
\includegraphics[width=.55\textwidth]{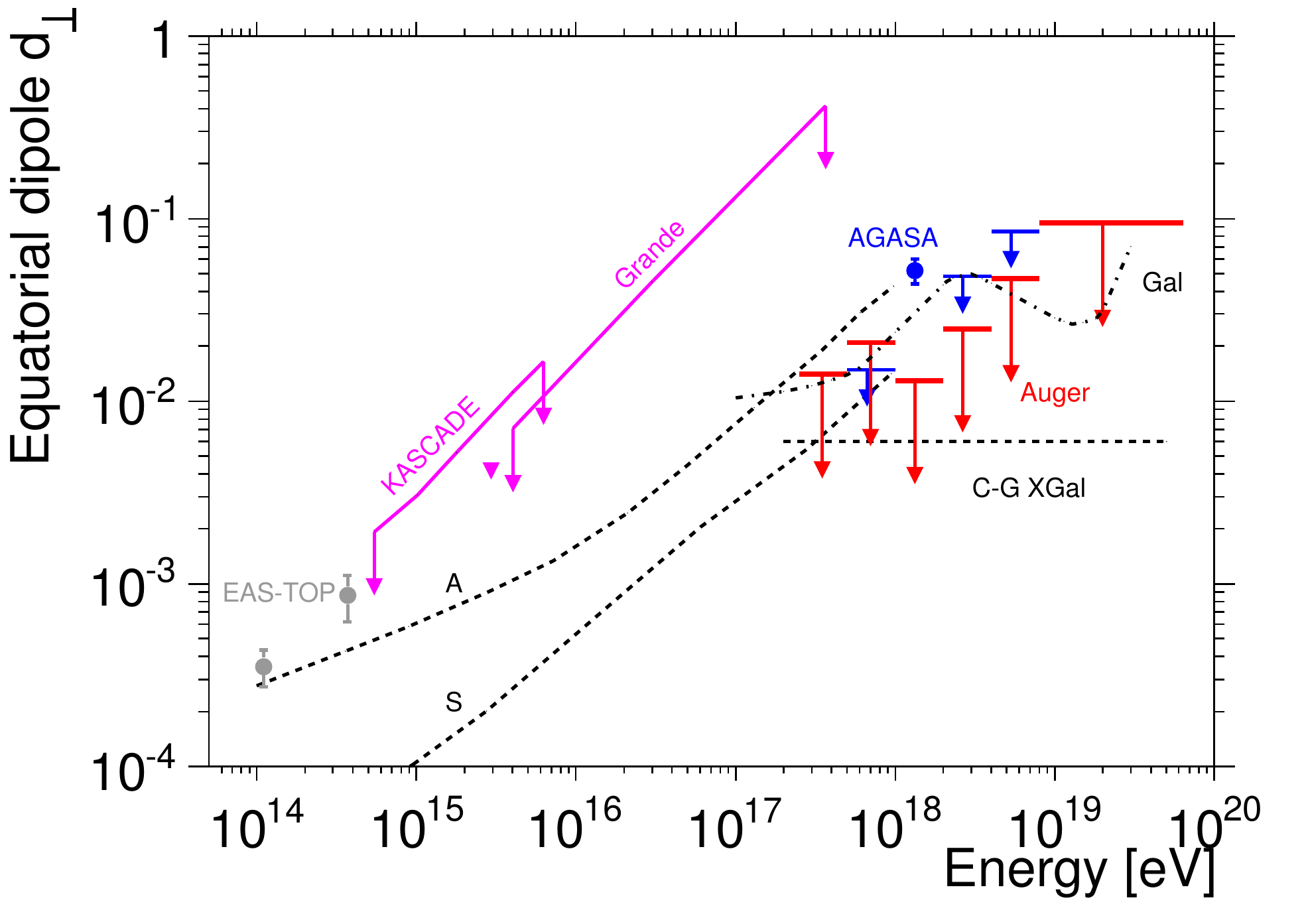}
\includegraphics[width=.45\textwidth]{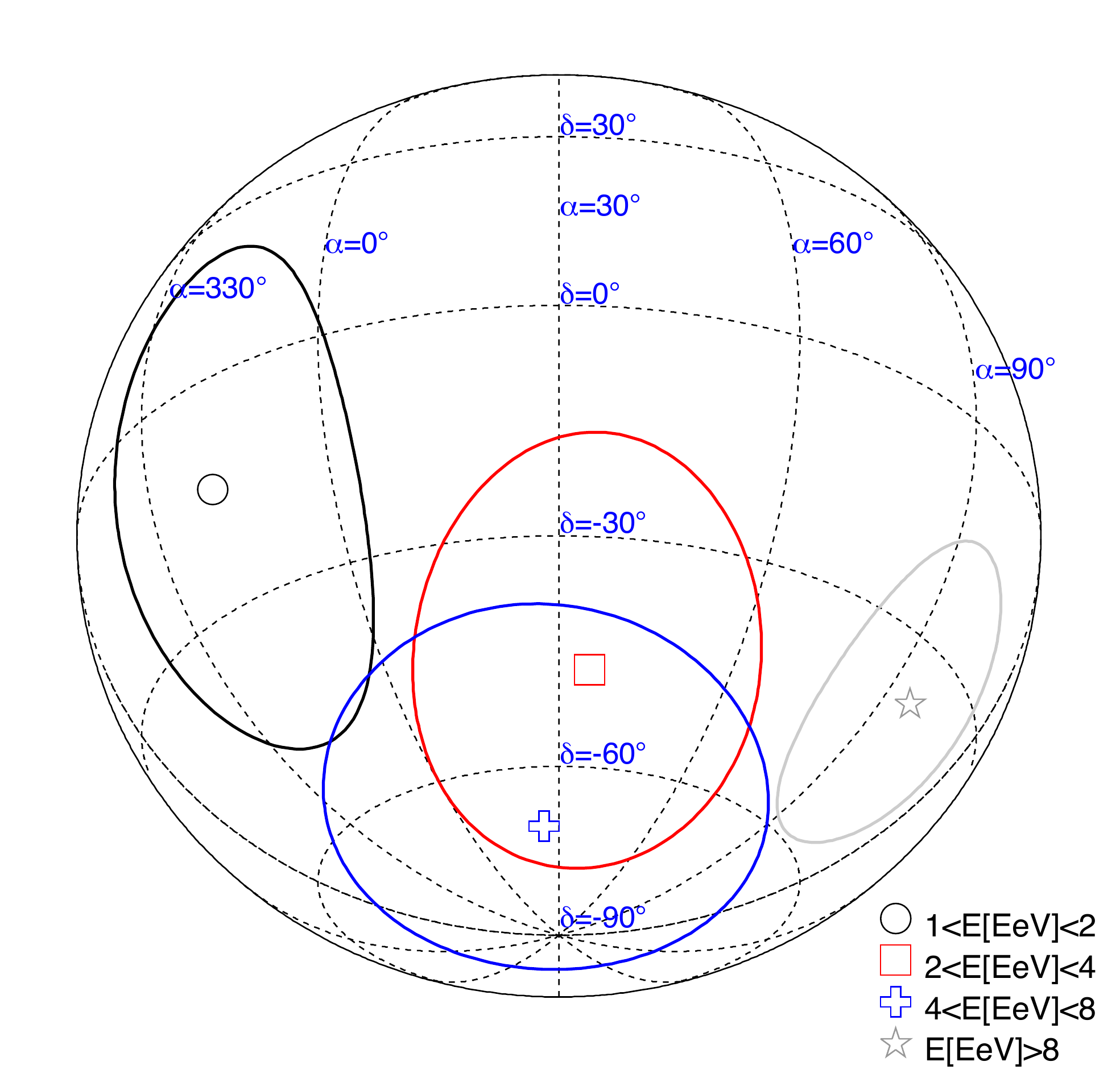}
\caption{
Left: Upper limits on the equatorial dipole component as a function of energy, from several experiments \cite{Abreu:2011ve}. Also shown are the predictions up to 1 EeV from two different Galactic magnetic field models with different symmetries (A and S), the predictions for a purely Galactic origin of UHECRs up to a few tens of $10^{19}$\,eV (Gal), and the expectations from the Compton-Getting effect for an EG component isotropic in the CMB rest frame (C-G Xgal) with references given in \cite{Abreu:2011ve}.
Right: Reconstructed declination and right-ascension of the dipole with corresponding uncertainties, as a function of the energy, in orthographic projection \cite{PAbreuetal:2012ve}.\label{fig:aniso}}
\end{figure}

The main obstacle in identifying Galactic CR-sources is the diffusion of CRs in the Galactic magnetic field (GMF), erasing directional information on the position of their sources. The GMF has a turbulent component that varies on scales between $l_{\rm min} \sim 1$~AU and $l_{\rm max}$ few to 200 pc. Since CRs scatter on inhomogeneities with variation scales comparable to their Larmor radius, the propagation of Galactic CRs in the GMF resembles a random walk and is well described by the diffusion approximation. Large scale anisotropies observed by the Tibet Air-Shower experiment \cite{Amenomori-06b} in the northern hemisphere for CRs at energies of a few to several hundred TeV and at angular scales of $60^\circ$ and above, thus came as a surprise. The data have been confirmed and complemented by Milagro \cite{Abdo-08} and more recently also by high statistics measurements of IceTop in the southern hemisphere \cite{Abbasi:2010jw,Abbasi:2012jv} (cf.\ Fig.\,\ref{fig:aniso}). Moreover, the structure changes with energy and appears to persist to beyond PeV energies. This anisotropy reveals a new feature of the Galactic cosmic-ray distribution, which must be incorporated into theories of the origin and propagation of cosmic rays. As was emphasized e.g.\ in \cite{Giacinti:2012hn,Aloisio:2012tx}, changes of the anisotropy patterns with energy can, in principle, be accounted for by specific distributions (in space and time) and individual source energy spectra of nearby recent SNRs.

Another long-standing problem is the high level of isotropy even at energies beyond $10^{18}$\,eV. For non-relativistic diffusive acceleration $\gamma_g = 2$ and the index of the observed spectrum, $\gamma_g + \mu = 2.7$, one derives $\mu = 0.7$. At very high energy this results in a too large anisotropy, $\delta(E) \propto D(E) \propto E^\mu$ and in a too small traversed grammage, $X_{\rm cr}(E) \propto 1/D(E)$, with the diffusion coefficient $D(E)$, and would contradict experimental data (cf.\ Fig.\ \ref{fig:aniso}).
However, $\delta(E) \propto D(E)$ refers again to an average of the anisotropy amplitude computed over many source realizations, i.e.\ over a continuum of sources and thus does not apply to concrete realizations of nearby SNRs and turbulent magnetic fields within the cosmic ray scattering length. Moreover, as pointed out in \cite{Giacinti:2012hn}, diffusion breaks down for propagation of CRs within our Galaxy at $E \ga 10^{17}$\,eV.
Recently, the Pierre Auger Observatory reported the first large scale anisotropy searches as a function of both right ascension and declination. Again, within the systematic uncertainties, no significant deviation from isotropy is revealed and the upper limits on dipole and quadrupole amplitudes challenge an origin from stationary galactic sources densely distributed in the galactic disk and emitting predominantly light particles in all directions. In Fig.\,\ref{fig:aniso} (right), the corresponding directions are shown in orthographic projection with the associated uncertainties, as a function of the energy. Both angles are expected to be randomly distributed in the case of independent samples whose parent distribution is isotropic. It is thus interesting to note that all reconstructed declinations are in the equatorial southern hemisphere, and to note also the intriguing smooth alignment of the phases in right ascension as a function of the energy.

Directional correlations of the most energetic CRs with nearby AGN observed by the Pierre Auger Observatory provided the first signature about anisotropies of the most energetic CRs and thereby about their EG origin \cite{Abraham-07e,Abraham-08a,Abreu-10}. Initially very strong, the fraction of Auger events above 55 EeV correlating within $3.1^\circ$ with a nearby ($z\le0.018$) AGN from the VCV-catalogue has stabilized at a level of $(33\pm5)$\,\% \cite{Kampert-ICRC}. With an accidental rate for an isotropic distribution of 21\,\%, this corresponds to a chance probability of less than 1\,\%. Recently, TA reported a correlation fraction of 44\,\% at an isotropic fraction of 24\,\% yielding a chance probability of about 2\,\% \cite{AbuZayyad:2012th}. Thus, the data are in perfect agreement with each other yielding a combined chance probability of observing such a correlation at the $10^{-3}$ level. However, more statistics is needed to consolidate the picture and to allow subdividing data sets in bins of related CR observables. The sky region around Cen A remains populated by a larger number of high energy events compared to the rest of the sky, with the largest departure from isotropy at $24^\circ$ around the center of Cen A with 19 events observed and 7.6 expected for isotropy, corresponding to a chance probability for this to occur at a level of 4\,\% \cite{Kampert-ICRC}. 

The Pierre Auger Collaboration has also performed a search for ultra-high neutrons from sources located within the Galaxy \cite{PAO:2012go}. Their mean path length is $9.2 \times E$\,kpc before decaying, where $E$ is the energy of the neutron in EeV. A stacking analysis was performed in the direction of bright Galactic gamma-ray sources:  the ones detected by Fermi-Lat instrument in the 100~MeV - 100~GeV range and the ones detected by H.E.S.S. in the range of 100~GeV - 100~TeV. Neither analysis provided evidence for significant excess and upper limits on the flux were derived for all directions within the Auger coverage. For directions along the Galactic plane for instance, the upper limits are below 0.024 km$^{-2}$yr$^{-1}$, 0.014 km$^{-2}$yr$^{-1}$
and 0.026 km$^{-2}$yr$^{-1}$ for energy bins of [1-2]~EeV, [2-3]~EeV and $E\geq1$~EeV, respectively.
For energies above 1~EeV, the 95 \% C.L.\ upper limit on the flux is 0.065 km$^{-2}$yr$^{-1}$, which corresponds to the energy flux of 0.13~EeV km$^{-2}$yr$^{-1}$ =  0.4~eV cm$^{-2}$s$^{-1}$ in the EeV range. Here a differential energy spectrum $E^{-2}$ was assumed~\cite{PAO:2012go}.
Those upper limits call into question the existence of persistent sources of EeV protons in the Galaxy. They also place useful constraints on the UHE emissions from the known Galactic sources of TeV gamma-rays. 

\section{Advances in Theory}
The observed knees may be interpreted as the end of the Galactic CR spectrum with their position defining the maximum acceleration energy. 
Although this picture may look very natural, the way to its adoption was not straight forward (see e.g.\ \cite{Aloisio:2012tx} for a recent review). Diffuse shock acceleration in SNR was discovered in the late 1970s but it took two decades more to realize that CR-streaming amplifies the magnetic fields upstream of the shock to create highly turbulent fields with strengths up to $\delta B \sim B \sim 10^{-4}$\,G (see e.g.\,\cite{lucek-00,bell-01}). The importance of non-linear magnetic fields amplification at SNR shocks, now also well accounted for by 3d MHD simulations, surely became one of the most discussed theory topics recently and there is general consensus that this is vital to reach the knee even on relatively short timescales on the order of tens to hundreds of years, i.e.\ much faster than it takes to reach the Sedov phase.

This drives the discussion to another important theoretical CR-problem. It is challenging to allow CRs to escape from SNR into the interstellar medium in large numbers without significant energy losses. This issue was first addressed by Ptuskin and co-workers \cite{Ptuskin:2006ew} where it was shown that only particles accelerated up to the maximum energy $E_{\rm max}$ can escape the acceleration region. In their original model, $E_{\rm max}$ reaches its highest value only at the beginning of the Sedov phase. The $E^{-2}$ CR energy spectrum in this model is then given by integrating the narrow peak of $E_{\rm max}(t)$ over time.

The diffusion of Galactic CRs close to their sources has been addressed recently by Kachelrie{\ss} and coworkers (see also \cite{Giacinti:2012hn}). Propagating individual CRs in purely isotropic turbulent magnetic fields with maximal scale of spatial variations $l_{\rm max}$, they find anisotropic CR diffusion at distances $r < l_{\rm max}$ from their sources. As a result, the CR densities around the sources become strongly irregular and show filamentary structures that could be probed by TeV $\gamma$-rays.

At the highest energies, Radio Galaxies (RG) remain being the most promising candidates for UHECR acceleration. An interesting argument linking UHECR sources to their luminosity at radio frequencies has been put forward in this context by Hardcastle \cite{Hardcastle:2010gm} and he concludes that RGs can accelerate protons to the highest observed energies in the lobes if a substantial amount of energy is in the turbulent component of the magnetic field, i.e.\ $B \ga B_{\rm equipart}$, and the Hillas criterion is met. In Cen A, existing observations do in fact constrain $B \ga B_{\rm equipart}$ for the kpc-scale jet. Moreover, if UHECRs are predominantly protons, then very few sources should contribute to the observed flux. These sources should be easy to identify in the radio and their UHECR spectrum should cut off steeply at the observed highest energies. In contrast, if the mass composition is heavy at the highest energies then many radio galaxies could contribute to the UHECR flux but due to the much stronger deflection only the nearby Radio Galaxy Cen A may be identifiable \cite{Hardcastle:2010gm}. Of course, such a conclusion depends very much on the strength of the EG magnetic fields and the maximum energy reached in the sources.


\section{New Projects and Outlook}

Motivated by the large body of important experimental findings and new insights, the field continues to evolve very dynamically with new projects being planned or existing ones to be upgraded. In the study of low energy CRs, {\bf AMS} is the by far most complex instrument in orbit, launched on May 16, 2011. It will measure light CR isotopes from about 500~MeV to 10~GeV and is hoped to improve our understanding of CR propagation in Galaxy. First data about the positron/electron ratio have been released recently \cite{Aguilar:2013hm} and confirm findings of PAMELA.
The {\bf CALET} (CALorimeteric Electron Telescope) project is a Japanese led international mission being developed as part of the utilization plan for the International Space Station (ISS) and aims at studying details of particle propagation in the Galaxy by a combination of energy spectrum measurements of electrons, protons and higher-charged nuclei. 
In Siberia, the German-Russian project {\bf HiScore} is planned to be constructed at the Tunka site. This project will use open Cherenkov counters for CR measurements around the knee and will be complemented by radio antennas to explore this new detection technology. {\bf HAWK} is being constructed in Mexico. Although its prime goal is the study of the $\gamma$-ray sky above 100~GeV, it will also contribute to measuring CR anisotropies at TeV-energies. {\bf LHAASO}, mostly driven by the Chinese community and much larger and more complex than HAWK, serves the same scientific goals. 

At the highest energies, {\bf Auger} and {\bf TA} plan upgrades in performance and size, respectively: Auger aims at improving the mass composition measurement and particle physics capabilities at the highest energies to answer the question about the origin of the flux suppression and TA aims at increasing their surface detector with a 2\,km grid up to 2800\,km$^2$.
Both collaborations have started to join efforts for a Next Generation Ground-based CR Observatory {\bf NGGO}, much larger than existing experiments and aiming at good energy and mass resolution and exploring particle physics aspects at the highest energies. Four proposed and planned space missions constitute the roadmap of the space oriented community: {\bf TUS, JEM-EUSO, KLPVE, and Super-EUSO} aim at contributing step-by-step to establish this challenging field of research. They will reach very large exposures aimed at seeing CR sources, which will be at the expense of energy resolution, composition measurements and particle physics capabilities. Given the resources of funding available in the next decade or two, it is unlikely that all of the above mentioned projects can be realized. Thus, priority should be given to complementarity rather than on duplication.

\begin{theacknowledgments}
Its a pleasure to thank the organizers of the Texas Symposium 2012 for inviting me to participate in this vibrant conference. I am also grateful for many stimulating discussions with colleagues from KASCADE-Grande, Auger and TA. Financial support by the German Ministry of Research and Education (Grants 05A11PX1 and 05A11PXA) and by the Helmholtz Alliance for Astroparticle Physics (HAP) is gratefully acknowledged.\end{theacknowledgments}



\end{document}